\documentclass[preprint,authoryear,12pt]{elsarticle}
\usepackage{graphics}
\usepackage{natbib}
\usepackage{pdflscape}

\journal{Advances in Space Research}

\begin{document}

\begin{frontmatter}

\title{O~{\small VI} absorption in the Milky Way along the Large Magellanic Cloud lines of sight}


\author{Rathin Sarma}
\address{Department of Physics, Hojai College, Hojai, India}

\author{Amit Pathak\corref{cor}}
\address{Department of Physics, Tezpur University, Tezpur 784 028, India}
\cortext[cor]{Corresponding author}
\ead{amit@tezu.ernet.in}

\author{Ananta C. Pradhan}
\address{Tata Institute of Fundamental Research, Homi Bhabha Road, Coalaba, Mumbai 400 005, India}

\author{Jayant Murthy}
\address{Indian Institute of Astrophysics, Koramangala, Bangalore 560 034, India}

\author{Jayanta K. Sarma}
\address{Department of Physics, Tezpur University, Tezpur 784 028, India}


\begin{abstract}
We have used \textit{Far Ultraviolet Spectroscopic Explorer (FUSE)} observations of the Large Magellanic Cloud (LMC) to determine the O~{\small VI} column densities in the Milky Way (MW) towards 6 LMC lines of sight. The mean column density of O~{\small VI} in the MW is found to be log N(O~{\small VI})=14.257$_{-0.084}^{+0.096}$. The results confirm the patchiness of O~{\small VI} absorption in the MW and the column densities are higher or comparable to the LMC.

\end{abstract}


\end{frontmatter}

\parindent=0.5 cm

\section{Introduction}

Interstellar medium (ISM) of galaxies provide crucial information about the processing of interstellar matter and the associated energy. The study of the ISM also adds to our knowledge on the formation of stars and stellar remnants. ISM is studied via the absorption and emission processes that are closely related with the radiation emitted by background or nearby stars. The ISM of galaxies show up in different electromagnetic bands with each of them giving information about different physical properties. While the X-ray is emitted by hot gas, UV bands directly tells us about several atomic species present in the ISM. Infrared is mostly thermal emission from interstellar dust which is transformation of the absorbed UV energy.

The O$^{+5}$ (O~{\small VI}) ion absorption line at 1032 \AA\ and 1038 \AA\ is an important tool to understand various physical processes and the life cycle of the warm-hot interface of the ISM of galaxies. The formation of O~{\small VI} ion in the ISM is through collisional ionization at temperatures of around 3$\times$10$^{5}$~K \citep{Indebetouw04, cox 2005}. Such temperatures are at the interface of warm (T $\sim 10^{4}$ K) and hot (T $>$ $10^{6}$ K) interstellar gas, thus, rendering the interstellar O~{\small VI} absorption and emission a sensitive gauge to monitor the interface matter, its structure and distribution. The detailed analysis of O~{\small VI} absorption helps in understanding the kinematics and velocity distribution associated with the gas, the total abundance and its comparison with other species and the processes leading to its formation in the ISM \citep{Wakker12}.

The {\it Copernicus} satellite \citep{Jenkins 1978a, Jenkins 1978b} and the {\it Hopkins Ultraviolet Telescope} \citep{Dixon et al. 1996} were the main instruments used for the study of the O~{\small VI} absorption with a limited spectral resolution. The launch of the {\it Far Ultraviolet Spectroscopic Explorer (FUSE)} \citep{Moos et al. 2000, Sahnow et al. 2000} in 1999 June has resulted in the inflow of quality data with high signal to noise. The {\it FUSE} has a high resolution of R$\sim$15,000-20,000 and a far-ultraviolet wavelength coverage of  905 -- 1187 \AA\ that fits very well for a detailed study of the O~{\small VI} spectra at 1032 \AA\ and 1038 \AA\ in the local universe. The study of O~{\small VI} absorption in the ISM of galaxies with the {\it FUSE} has given us information about the formation and distribution of O~{\small VI} in the Milky Way (MW) and the Magellanic Clouds that adds to our knowledge of varying ISM conditions in environments of different metallicities \citep{Savage00, Wakker03, Oegerle05, Savage06, Welsh08}. Apart from absorption studies, the {\it FUSE} has also been used to observe the O~{\small VI} spectra in emission from the diffuse ISM in the MW \citep{Shelton01, Dixon06, Dixon08} and superbubbles (SBs) in the Large Magellanic Cloud (LMC) \citep{Sankrit07}. Such studies have added to our understanding of the ISM and the physical processes in different interstellar environments. Apart from this, the O~{\small VI} absorption at low redshifts trace the warm-hot intergalactic medium (WHIM) and is an important contributer to model the cosmological problem of missing baryons \citep{Tepper-Garcia11}.

We reported a survey of O~{\small VI} absorption in the LMC using the {\it FUSE} observational data \citep{Pathak et al.11}. Towards all the lines of sight, these observations also have a MW component that remains  to be studied. Here we report the {\it FUSE} observations of interstellar O~{\small VI} in the MW as observed towards 6 lines of sight along the LMC [(l, b) = 280$^\circ$, -32$^\circ$]. These observations are new and have not been reported earlier as per our knowledge. Considering the fact that analysis of only six sightlines is discussed, a comparison of the results with the O~{\small VI} absorption in the LMC and the Small Magellanic Cloud (SMC) may not be relevant. Nevertheless, the results are an indication of the variation of the amount of O~{\small VI} in varying metallicity environments. The results may also be used for mapping the O~{\small VI} absorption in the MW and study the small scale variation along different sightlines. In Section 2 we explain the observations and data analysis. We discuss the column densities in Section 3 and summarize our results in Section 4.

\section{Observations and data analysis}

The {\it FUSE} \citep{Moos et al. 2000, Sahnow et al. 2000} observes through three apertures the LWRS, the MDRS and the HIRS simultaneously. Here we report data acquired with the LWRS (30$\times$30 arcsec$^{2}$) and the MDRS (4$\times$20 arcsec$^{2}$) apertures. The fully calibrated {\it FUSE} spectra were downloaded from the Multimission Archive at STScI (MAST) processed by the latest {\it FUSE} data reduction pipeline CALFUSE version 3.2 \citep{Dixon07}. Though, the {\it FUSE} has more than 600 observations in and around the LMC, here we report analysis of six lines of sight (Table \ref{Tab1}). The six lines of sight have been chosen as they show resolved O{\small VI} absorption for the LMC and MW velocities. A detailed analysis of the MW O~{\small VI} absorption for the complete {\it FUSE} LMC observational data set is ongoing.

\begin{figure}
\includegraphics{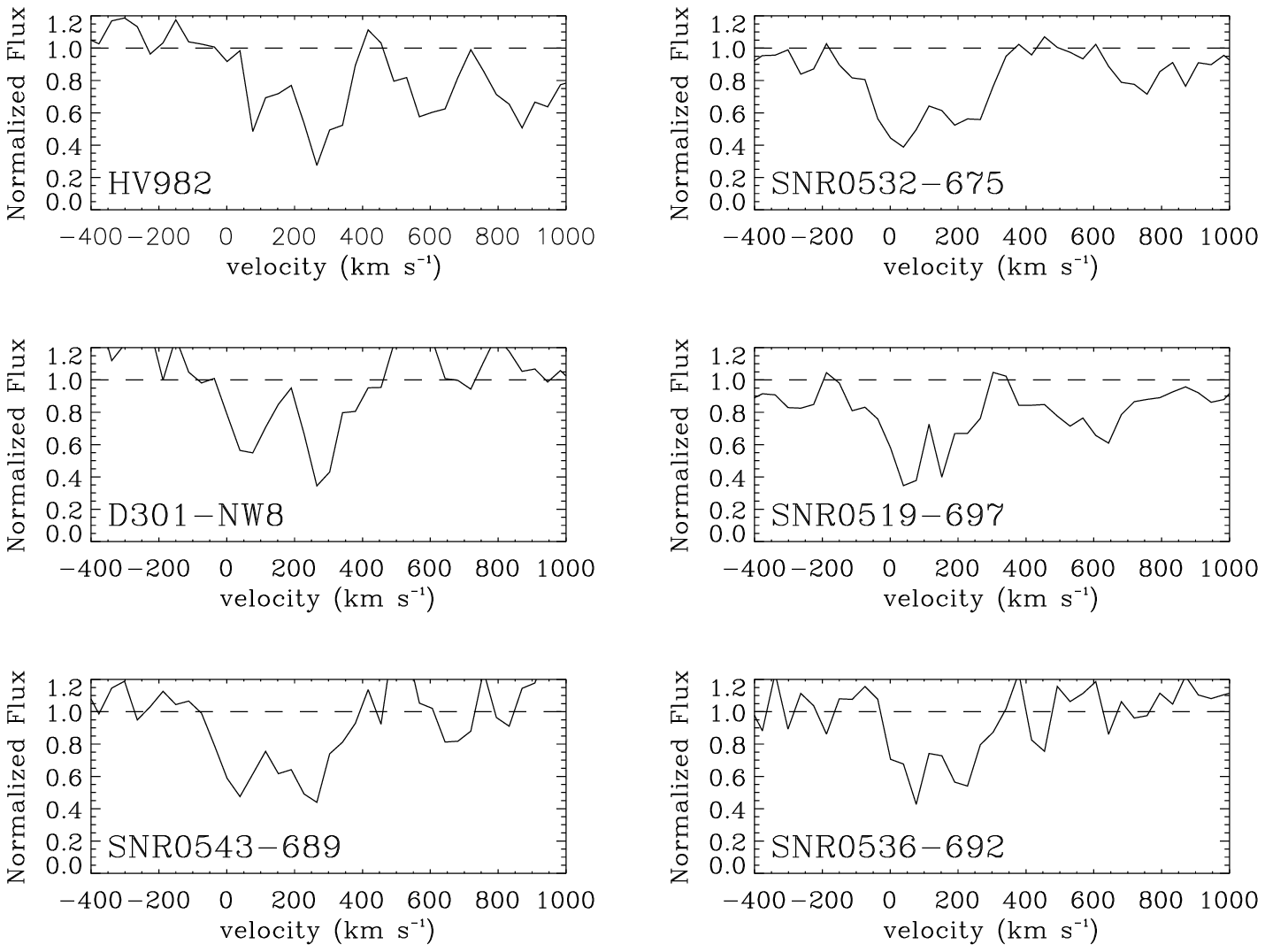}
\caption{O~{\small VI} absorption profiles for 6 lines of sight in the MW. Table 1 gives the details of each sightline.} 
\label{Fig1}
\end{figure}

In order to have a higher signal-to-noise, we have downgraded all the 6 spectra to 35 km~s$^{-1}$. Following the spectral analysis procedure of \citet{Howk02} and \citet{Sembach92}, the local stellar continuum was estimated for all the targets and were fitted by a Legendre polynomial fit of low order \citep{Pathak et al.11}. Out of several test continua, the best was selected and the uncertainties involved was added up in estimating the errors in the measurement of the O~{\small VI} column densities. It should be noted that we have not corrected for any stellar wind absorption features. The normalized spectra in the vicinity of O~{\small VI} absorption are presented in Fig. \ref{Fig1} and the equivalent widths and the column densities are presented in Table \ref{Tab2}.

The O~{\small VI} absorption may be contaminated by absorption from molecular hydrogen. Though the contamination by molecular hydrogen absorption in the LMC velocity range is minimal \citep{Pathak et al.11}, we  have accounted for it in the presented analysis of the MW measurements. We have theoretically modeled the contamination by H$_2$ and subtracted it from the O~{\small VI} column density measurements for the MW velocities.

\section{O~{\small VI} column densities}
We use the apparent optical depth technique \citep{Savage91, Sembach92, Howk02} to measure the equivalent width and column density of the O~{\small VI} absorption. This technique uses an apparent optical depth ($\tau_a$) in terms of velocity,
\begin{equation}
\tau_a(v) = ln[I_{o}(v)/I_{obs}(v)],
\end{equation}
where, $I_o$ is the estimated continuum intensity and $I_{obs}$ is the intensity of the absorption line in terms of velocity. Considering that the resolution of the {\it FUSE} is high compared to the FWHM of the absorption line, the apparent optical depth is a good approximation to the true optical depth. The apparent column density ($N_a(v)$ [atoms cm$^{-2}$ (km~s$^{-1}$)$^{-1}$]) may be calculated by the following relation
\begin{equation}
N_a(v) = \frac{m_e c \tau_a(v)}{\pi e^2 f \lambda} = 3.768 \times 10^{14} \frac{\tau_a(v)}{f \lambda},
\end{equation}
where $m_e$ is the mass of the electron, $c$ is the speed of light, $e$ is the electronic charge, $\lambda$ is the wavelength (in \AA) and {\it f} is the oscillator strength of the atomic species (for O~{\small VI}, {\it f} value of 0.1325 has been adopted from \citet{Yan98}). The 1032 \AA\ O~{\small VI} profile is broad and is completely resolved by {\it FUSE}. However, the weaker of the O~{\small VI} doublet at 1037.6 \AA\ is found to be inseparable from the CII* and H$_2$ absorptions.

The metallicity of the MW is higher than that of the LMC. This directly implies that the O~{\small VI} abundance in the MW should be higher compared to that of the LMC. Table \ref{Tab2} presents a comparison of the equivalent widths and the column densities in the two galaxies. We find that the O~{\small VI} column densities are indeed higher but also has a considerably lower value for the target HV 982. Apart from the 1$\sigma$ uncertainties provided in Table 2, the major source of error is the overlap of the MW O~{\small VI} absorption profile with the LMC O~{\small VI} absorption. Since, the O~{\small VI} absorption is very patchy in nature and the O~{\small VI} abundance depends on local ISM conditions, it is extremely difficult to compare the O~{\small VI} column densities of the MW and the LMC. Our other limitation here is that we have analysed only six sightlines which definitely are not enough to yield a conclusion. Our future aim is to analyse a bigger data set and compare the abundance of O~{\small VI} in the MW, the LMC and the SMC and see if it follows the metallicity of the ISM in the three galaxies.

\citet{Meaburn84} have observed neutral gas in the MW along the LMC and have identified velocity components associated with the Galactic disk and the halo of the Galaxy. This has further been confirmed by components of metal absorption lines \citep{Savage81}. O~{\small VI} absorption in the MW for 12 lines of sight along the LMC have been presented by \citet{Howk02a}. \citet{Howk02a} have discussed the contribution of high velocity clouds (HVCs) to the MW O~{\small VI} absorption. The results presented here give the MW column densities of 6 additional sightlines (Table \ref{Tab2}). Using these results, we try to study the variation in the O~{\small VI} column density. The second column of Table \ref{Tab2} gives the Galactic latitudes and longitudes for the {\it FUSE} pointings. The third column of Table \ref{Tab2} gives the distance of each source from the HV 982 (first source). The distances have been calculated assuming O~{\small VI} absorption by HVCs located in the halo of the Galaxy towards the LMC at a distance of 40 kpc. Table \ref{Tab2} clearly indicates the variation in the amount of O~{\small VI} at small scales (from a minimum of 0.4 kpc to a maximum of 1.8 kpc) and confirms the patchiness.

\section{Summary and conclusion}

We have presented O~{\small VI} column density measurements for the MW along six sightlines towards the LMC using {\it FUSE} spectroscopic data. The highest column density measured for the MW is log N(O{\small VI}) = 14.508~atoms~cm~$^{-2}$ and the minimum value is log N(O{\small VI}) = 13.911~atoms~cm~$^{-2}$. We find that overall, the MW column densities are higher or comparable to that of the LMC except for the source HV 982. The number of sightlines studied are too few to give a concrete inference from the comparison between the two galaxies, still the indications are in the right direction. We have also confirmed the patchiness of O~{\small VI} in the MW. This is in accordance with earlier measurements of O~{\small VI} absorption for the Galaxy. This work is part of an effort to carry out a survey of the O~{\small VI} absorption in the MW and the Magellanic Clouds. A bigger data set will allow us to do a detailed analysis of the origin of patchiness in O~{\small VI} absorption.

\section{Acknowledgments}

AP acknowledges seed money grant from Tezpur University and Visiting Associateship of the Inter-University Centre for Astronomy and Astrophysics, Pune.

\clearpage

\clearpage

\begin{landscape}
\begin{table}
\scriptsize
\begin{center}
\caption{Log of \textit{FUSE} observations for the 6 targets in the Milky Way.}
\label{Tab1}
\begin{tabular}{lccccccc}

\hline
\hline
\textit{FUSE ID} &   Object name & \textit{FUSE} aperture &  RA & Dec.& Spectral type$^1$& V(mag)$^1$ & Reference$^2$ \\
&&&($^{h m s}$)&($^{\circ}$ $^{\prime}$ $^{\prime \prime}$)&&&\\
\hline

C1030201 & HV982 & LWRS & 05 27 27.40 & -67 11 55.4 & B0-B2V-IV & 14.6 & 1 \\
D0981501 & D301-NW8 & MDRS & 05 43 15.96 & -67 49 51.0 & D301-NW8 & 14.37 & 2\\
D9044701 & SNR0519-697 & LWRS & 05 18 44.20 & -69 39 12.4 & Supernova remnant&\\
D9043101 & SNR0536-692 & LWRS & 05 36 07.70 & -69 11 52.6 & Supernova remnant&\\
D9042801 & SNR0543-689 & LWRS & 05 43 07.20 & -68 58 52.0 & Supernova remnant & \\
D9042002 & SNR0532-675 & LWRS & 05 32 23.00 & -67 31 02.0 & Supernova remnant&\\
\hline
\end{tabular}
\end{center}
1 - Spectral type and V (mag) is for the background object in the LMC.\\
2 - References: (1)\citep{Blair et al. 2009}; (2)\citep{Massey 2002}
\end{table}
\end{landscape}

\clearpage

\begin{landscape}
\begin{table}
\scriptsize
\begin{center}
\caption{O~{\small VI} column densities, equivalent widths and the corresponding velocity limits in the Milky Way and the LMC.}
\label{Tab2}
\begin{tabular}{lcccccccc}

\hline
\hline
Target name & Galactic (l, b) & distance$^1$ & \multicolumn{3}{c}{Milky Way}&  \multicolumn{3}{c}{LMC} \\
&&(kpc)& EW&log N(O~{\small VI})&Integration limit&EW&log N(O{\small VI})&Integration limit\\
&&&(m\AA)&(dex)&(km~s$^{-1}$)&(m\AA)&(dex)&(km~s$^{-1}$)\\
\hline

HV982       &277.39, -32.95 &0 &117.5$\pm$54.8 &13.911$_{-0.07}^{+0.08}$ &40, 175 &208$\pm$58 &14.33$_{-0.17}^{+0.12}$ &175, 360\\  
D301-NW8    &277.93, -31.37 &1.14 &204.2$\pm$35.5&14.288$_{-0.10}^{+0.10}$ &-30, 175  &228$\pm$30 &14.42$_{-0.03}^{+0.02}$ &175, 365\\
SNR0519-697 &280.44, -33.32 &1.80 &339.0$\pm$25.5&14.436$_{-0.11}^{+0.13}$ &2, 160    &97$\pm$4 &13.97$_{-0.08}^{+0.07}$ &160, 300\\
SNR0536-692 &279.61, -31.89 &1.50 &139.7$\pm$39.7&14.156$_{-0.07}^{+0.07}$ &38, 165  & 116$\pm$7& 14.03$_{-0.07}^{+0.05}$ &165, 320\\
SNR0543-689 &279.28, -31.29 &1.61 &238.8$\pm$8.47& 14.241$_{-0.09}^{+0.10}$&0, 160   &186$\pm$34 &14.27$_{-0.12}^{+0.10}$ &160, 360\\
SNR0532-675 &277.69, -32.44 &0.40 &322.0$\pm$61.4& 14.508$_{-0.05}^{+0.09}$&-70, 165 &147$\pm$9 &14.16$_{-0.08}^{+0.07}$ &165, 345\\               
\hline
\end{tabular}
\end{center}
1 - All the distances have been calculated from HV982 assuming the distance of the absorbing cloud to be 40 kpc from the Sun.\\
Notes. The LMC results have been taken from \citet{Pathak et al.11}.\\
The errors in the equivalent widths and column densities are 1$\sigma$ error estimates.

\end{table}
\end{landscape}

\end{document}